\documentclass{PoS_arXiv}

\usepackage{amsmath,amssymb,amsfonts,graphicx,cite,psfrag}

\input paperdef
\graphicspath{{figs/}}

\title{Tools for Charged Higgs Bosons}

\ShortTitle{Tools for Charged Higgs Bosons}

\author{\speaker{Sven Heinemeyer}\\ 
        Instituto de Fisica de Cantabria (CSIC--UC), Santander, Spain\\
        E-mail: \email{Sven.Heinemeyer@cern.ch}}

\abstract{
In order to identify the Higgs sector realized in nature, the
predictions for Higgs boson masses, production cross sections and decay
widths have to be compared with experimental results.
We give a brief overview about computer codes for the evaluation of the
properties of charged Higgs bosons, mostly focusing on the case of the
Minimal Supersymmetric Standard Model (MSSM). We briefly review the
relevance of the various contributions to the charged MSSM Higgs boson
mass arising at the one-loop level.
}

\FullConference{Prospects for Charged Higgs Discovery at Colliders\\
		 September 16-19 2008\\
		 Uppsala, Sweden}

\begin{document}


\section{Introdcution}

Identifying the mechanism of electroweak symmetry
breaking will be one of the main goals of the LHC. 
Many possibilities have been studied in the past, of which 
the most popular ones are the Higgs mechanism within the Standard Model
(SM)~\cite{sm}, the Two Higgs Doublet Model (THDM) \cite{thdm}, 
and within supersymmetric (SUSY) models~\cite{mssm}, which naturally
contain at least two Higgs doublets.
While the SM contains only one physical (neutral) Higgs boson, models
with two Higgs doublets contain, besides three neutral Higgs bosons,
at least two charged Higgs bosons, $H^\pm$. 
Any evidence for a charged Higgs boson would be an unambiguous sign of
an extended Higgs sector, i.e.\ of physics beyond the SM.

In order to identify the Higgs sector realized in nature, the
predictions for couplings to SM fermions and bosons as well as
self-couplings have to be compared with the experimental results. 
To perform this task precise theory predictions are needed for Higgs
production cross sections and decay widths (or at best for complete
processes). Here we give a brief overview about this kind of codes
concerning the charged Higgs bosons. A recent overview about codes and
tools for SUSY can be found in \citere{AllanachTools}.

\smallskip
Within the THDM the Higgs potential contains 14 independent parameters,
and the mass of the charged Higgs boson, $\MHp$, is usually treated as a free
parameter. Consequently, no code exists predicting $\MHp$. However,
tree-level production cross sections and decay width can be evaluated
with codes like {\tt MadGraph}~\cite{madgraph}, 
{\tt Pythia}~\cite{pythia}, {\tt Sherpa}~\cite{sherpa} or
{\tt Herwig}~\cite{herwig}. We are not aware of codes containing
higher-order corrections to charged Higgs boson processes. However, this
could be done in principle with 
{\tt FeynArts}/{\tt FormCalc}~\cite{feynarts,formcalc}, for which the
THDM model file is available.

Within SUSY many results for the charged Higgs boson beyond the
tree-level are available in the Minimal Supersymmetric Standard Model
(MSSM), see below. The only code that contains higher-order corrections
for a charged Higgs boson beyond the MSSM is 
{\tt NMSSMTools}~\cite{nmssmtools}, providing masses, production cross
sections and decay widths in the NMSSM, i.e.\ the MSSM with an
additional Higgs singlet. In the following we will focus
on the MSSM.


\section{The charged Higgs in the MSSM}

The Higgs sector of the MSSM
contains two Higgs doublets, leading to five
physical Higgs bosons. At tree-level these are the light and heavy
$\cp$-even $h$~and $H$, the $\cp$-odd $A$ and the charged $H^\pm$. At
lowest order the Higgs sector can be described besides the SM
parameters by two additional independent
parameters, chosen to be the mass of the $A$~boson, $\MA$ 
(in the case of vanishing complex phases), and 
$\tb \equiv v_2/v_1$, the ratio of the two vacuum expectation values. 
Accordingly, all other masses and couplings can be predicted at
tree-level, e.g.\ the charged Higgs boson mass
\BE
\label{MHptree}
\mHp^2 = \MA^2 + \MW^2~.
\EE
$M_{Z,W}$ denote the masses of the $Z$~and $W$~boson,
respectively. 
This tree-level relation receives higher-order corrections, where the loop
corrected charged Higgs-boson mass is denoted as $\MHp$.

The charged Higgs bosons of the MSSM (or a more general Two Higgs
Doublet Model) have been searched at LEP and the Tevatron, and will be
searched for at the LHC and the ILC. The LEP 
searches~\cite{ADLOchargedHiggs,LEPchargedHiggsPrel}, 
yielded a bound of
$\MHp \gsim 80 \gev$~\cite{LEPchargedHiggs}.
The Tevatron is especially sensitive at low $\MA$ and large $\tb$,
extending the LEP bounds in this region of parameter
space~\cite{Tevcharged}. Also at the LHC the prospects for charged Higgs
boson searches are best at large $\tb$, reaching up to 
$\MA \lsim 800 \gev$~\cite{lhctdrs,benchmark3,cmsHiggs2}. At the
ILC, if the charged Higgs is in the kinematical reach,
a high-precision determination of the
charged Higgs boson properties will be
possible~\cite{Snowmass05Higgs,MHpLHCILCnewer}.

Many computer codes exists to evaluate various properties of the
(charged) Higgs boson in the MSSM:
\begin{itemize}

\item 
RGE running from a high-energy scale to the electroweak (EW) scale
for SUSY parameter is provided, for instance, by
{\tt SoftSusy}~\cite{softsusy},
{\tt Spheno}~\cite{spheno},
{\tt Suspect}~\cite{suspect} or
{\tt Isajet}~\cite{isajet}.
Subsequently they evaluate $\MHp$ including (some) one-loop corrections.

\item 
Three codes exist for the calculation of $\MHp$ and the various decay
widths, 
{\tt FeynHiggs}~\cite{feynhiggs,mhiggslong,mhiggsAEC,mhcMSSMlong,mhcMSSM2L}, 
{\tt CPsuperH}~\cite{cpsh}
and {\tt Hdecay}~\cite{hdecay}. 
More details on the evaluations can be found below.

\item
Calculations for the charged Higgs production at the LHC have been
performed in\\ \citeres{HpmXSa,HpmXSb} (see also \citere{HpmXS4f}), 
for the production in
association with a $W$~boson in \citere{ppHW} and for the production of
a $H^\pm H^\mp$ pair in \citere{ppHpHm}. However, no dedicated code
exists. The calculation of \citeres{HpmXSa,HpmXSb} has recently been
implemented into {\tt Prospino}~\cite{prospino} and is also included in 
{\tt FeynHiggs}~\cite{feynhiggs,mhiggslong,mhiggsAEC,mhcMSSMlong,mhcMSSM2L}
(including the $\db$ corrections, see below).
More details on recent higher-order corrections can be found
in \citere{HpmXSrev}. 
For the $H^\pm$ production at the ILC we are only aware of one code for
the calculation of $e^+e^- \to W^\pm H^\mp$~\cite{eeHW}.

\item
Event generators can produce events with a charged MSSM Higgs boson:
{\tt Pythia}~\cite{pythia}, {\tt Sherpa}~\cite{sherpa} or
{\tt Herwig}~\cite{herwig}. They contain only relatively few corrections
to the Higgs boson mass or decay widths. However, they can be linked to
the dedicated spectrum and decay calculators via the SUSY Les Houches
Accord~\cite{slha}. 

\item
Indirect constraints on the mass and couplings of the charged MSSM Higgs
boson can be set via (e.g.) $B$-physics observables. One dedicated code
for this is {\tt SuperIso}~\cite{superiso}, another one is described
in \citere{IsidoriParadisi}. Other codes contain some $B$-physics
observables as an additional check on the MSSM parameter space. Examples
are {\tt MicrOMEGAs}~\cite{micromegas}, {\tt CPsuperH}~\cite{cpsh} or
{\tt FeynHiggs}~\cite{feynhiggs,mhiggslong,mhiggsAEC,mhcMSSMlong,mhcMSSM2L}.
LEP and Tevatron bounds on the charged Higgs boson searches are
currently implemented into {\tt HiggsBounds}~\cite{higgsbounds}.

\end{itemize}

A ``comparison'' of the dedicated mass and decay width calculators has
to be split up into the case of the MSSM with real parameters (rMSSM)
and the MSSM with complex parameters (cMSSM). {\tt Hdecay} is purely for
real parameters. {\tt CPsuperH} and {\tt FeynHiggs} can handle also
parameters with complex phases (for instance 
$\At$, $\Ab$, $M_3$, $\mu$, \ldots).
The following corrections to $\MHp$ are implemented into the three codes
(we do not go into details concerning the attempt to capture three-loop
corrections): 

\begin{itemize}
\item[(i)] 
{{\tt Hdecay}: $\MHp$ in the rMSSM:}\\
$\MA$ is input parameter, and the higher-order corrections to $\MHp$ are
evaluated. The calculation of \citere{mhiggsRGE} (including corrections up to
the two-loop 
level) has been exteded to the complete Higgs potential and thus to $\MHp$,
including the $\db$ corrections.

\item[(ii)]
{{\tt CPsuperH}: $\MHp$ in the cMSSM:}\\
$\MHp$ is an input parameter, and the higher-order corrections appear
as shifts to the neutral Higgs boson masses. 
A large set of one-loop corrections are supplemented by two-loop
contributions obtained by RGE improvements, see \citere{mhiggsCPXRG1}
and references therein. The complex $\db$ corrections (see below) are
included. 

\item[(iii)]
{{\tt FeynHiggs}: $\MHp$ in the rMSSM:}\\
$\MA$ is an input parameter, and the higher-order corrections to $\MHp$
are evaluated. They comprise the full set of one-loop
contributions~\cite{mhcMSSMlong} and various two-loop corrections as
summarized in \citere{mhiggsAEC}.\\
{{\tt FeynHiggs}: $\MHp$ in the cMSSM:}\\
$\MHp$ is an input parameter, and the higher-order corrections appear
as shifts to the neutral Higgs boson masses. The calculation comprises
the full set of one-loop contributions~\cite{mhcMSSMlong} and
the \order{\alt\als} corrections as given in \citere{mhcMSSM2L}.

\end{itemize}

An overview about the various charged MSSM Higgs boson properties
implemented into {\tt FeynHiggs}, {\tt CPsuperH} and {\tt Hdecay} is
shown in \refta{tab:widthcalc}%
\footnote{
The Higgs-self couplings and the $\br(t \to H^+ b)$ are internally 
calculated in {\tt Hdecay}, but are so far not part of the
output~\cite{spira}. 
}%
.~The evaluation of the charged Higgs
decay contains in all three codes the contribution coming from $\db$
(see below). However, the effect of complex phases in the various 
calculations is taken into account only 
in {\tt FeynHiggs} and {\tt CPsuperH}. The Higgs self-couplings are
understood to contain an $H^+H^-$ pair. The LHC production cross section
calculation in in {\tt FeynHiggs} is based on \citeres{HpmXSa,HpmXSb},
supplemented by the $\db$ corrections.
We also list $\br(t \to H^+ \bar b)$, which be used to evaluate the
charged Higgs production cross section for $\MHp < \mt$ at the Tevatron
or the LHC.

\begin{table}[htb!]
\renewcommand{\arraystretch}{1.2}
\BC
\begin{tabular}{|c||c|c|c|} 
\cline{2-4} \multicolumn{1}{c||}{}
& {\tt FeynHiggs} & {\tt CPsuperH} & {\tt Hdecay} \\ 
\hline\hline
$\Ga_{\rm tot}$ & $\surd$ & $\surd$ & $\surd$ \\
$\br(H^+ \to f^{(*)} \bar f^\prime)$ & $\surd$ & $\surd$ & $\surd$ \\
$\br(H^+ \to h_i W^{(*)})$ & $\surd$ & $\surd$ & $\surd$ \\
$\br(H^+ \to \sfi \tilde{f}_j^\prime)$ & $\surd$ & $\surd$ & $\surd$ \\
$\br(H^+ \to \neu{i}\chap{j})$ & $\surd$ & $\surd$ & $\surd$ \\
\hline
Higgs triple self-couplings & $\surd$ & $\surd$ &         \\
Higgs quartic self-couplings &         & $\surd$ &         \\
\hline
$\si(pp \to H^+ + X)$ at the LHC & $\surd$ &         &         \\
$\br(t \to H^+ \bar b)$ & $\surd$ & $\surd$ &         \\
\hline
\end{tabular}
\EC
\vspace{-1em}
\caption{Overview about the various charged MSSM Higgs boson properties
implemented into {\tt FeynHiggs}, {\tt CPsuperH} and {\tt Hdecay}. 
$f$ denotes a SM fermion, $h_i$ ($i = 1, 2, 3$) are the three neutral
Higgs bosons, $\sfi$ denotes a scalar fermion, and $\neu{i}$, $\cha{k}$
are the neutralinos and charginos, respectively.
The evaluation of the charged Higgs
decay contains in all three codes the contribution coming from $\db$.
The effect of complex phases in the various calculations is taken into
account only  in {\tt FeynHiggs} and {\tt CPsuperH}. 
Higgs self-couplings are
understood to contain an $H^+H^-$ pair. The LHC production cross section
calculation in in {\tt FeynHiggs} is based on \citeres{HpmXSa,HpmXSb},
supplemented by the $\db$ corrections. 
The $\br(t \to H^+ \bar b)$ can be used to evaluate the charged Higgs
production cross section for $\MHp < \mt$.
}
\label{tab:widthcalc}
\renewcommand{\arraystretch}{1.0}
\end{table}


\section{Relevance of higher-order corrections to \boldmath{$\MHp$}}

\subsection{The higher-order corrections}

As mentioned above, 
the charged Higgs boson mass, see \refeq{MHptree}, receives higher-order
corrections, which are (at different levels of sophistication)
implemented into public codes.
In a first step in \citere{mhp1lA} leading corrections to the sum rule
given in  \refeq{MHptree} have been evaluated. 
Then one-loop corrections from $t/b$ and $\Stop/\Sbot$~loops have been
derived in \citeres{mhp1lB,mhp1lC}. These were extended to the one-loop
leading logarithmic terms from all sectors of the MSSM in
\citere{mhp1lD}. The 
first full one-loop calculation in the Feynman-diagrammatic (FD)
approach has been performed in \citere{mhp1lE}, and re-evaluated more
recently in \citeres{markusPhD,mhcMSSMlong}. 
Within the FD approach the leading \order{\alt \als} two-loop terms 
have recently been obtained in \citere{mhcMSSM2L,chargedHiggs2L}.
In the following we will focus on the one-loop corrections.

Within the FD approach the charged Higgs boson pole mass, $\MHp^2$, is
obtained at the one-loop level by solving the equation
\begin{align}
p^2 - \mHp^2 + \hSi_{H^+H^-}^{(1)}(p^2) = 0~.
\end{align}
$\hSi_{H^+H^-}^{(1)}(p^2)$ denotes the renormalized one-loop charged
Higgs boson self-energy. Details about the calculation and the full one-loop
evaluation can be found in \citeres{mhcMSSMlong,markusPhD}. 

Another class of higher-order corrections that is formally of two-loop
order is normally included already into the one-loop result.
These corrections originate from the contributions to Higgs boson
self-energies from the bottom/sbottom sector enhanced by $\mu\,\tb$. Large
higher-order effects can in particular occur in the relation between
the bottom-quark mass and the bottom Yukawa coupling at large $\tb$.
Because the $\tb$-enhanced
contributions can be numerically relevant, an accurate determination
of $h_b$ from the experimental value of the bottom mass requires a
resummation of such effects to all orders in the perturbative
expansion, as described in \citeres{deltamb2,deltamb2b}.
Effectively the bottom Yukawa coupling is given by
\begin{align}
\frac{g}{2\MW} \frac{\mbms}{1 + \db} \quad ~{\rm with}~ \quad
\mbms \; \equiv \; \mbms^{\DRbar}(Q) 
      \; = \; \mbms^{\MSbar}(Q) \KL 1 + \frac{4\,\als}{3\,\pi} \KR~.
\label{mbdrbar}
\end{align}
$\mbms$ denotes the running \DRbar\ bottom
quark mass at the mass scale~$Q$, where $\mbms^{\MSbar}$ includes
the SM QCD corrections. 
The leading one-loop contribution
$\db$ in the limit of $\msusy \gg \mt$ and $\tb \gg 1$ takes the simple
form~\cite{deltamb1}
\begin{align}
\db &= \frac{2\als}{3\,\pi} \, M_3^* \, \mu^* \, \tb \,
                    \times \, I(\msbe, \msbz, |M_3|) 
     + \frac{\alt}{4\,\pi} \, \At^* \, \mu^* \, \tb \,
                    \times \, I(\mste, \mstz, |\mu|) ~, 
\label{def:db}
\end{align}
where the function $I$ arises from the one-loop vertex diagram and
scales as\\
$I(a, b, c) \sim 1/\mbox{max}(a^2, b^2, c^2)$.
Here $\mste, \mstz$ and $\msbe, \msbz$ denote the two scalar top and
bottom masses, respectively. $|M_3|$ is the gluino mass parameter, $\mu$
is the Higgs mixing parameter, and $\At$ denotes the trilinear Higgs-stop
coupling. 
This type of higher-order corrections being formally beyond the one-loop
level are included into the Yukawa couplings appearing in the one-loop
result. 


\subsection{Numerical analysis}

The higher-order
corrected Higgs-boson sector has been evaluated with the help of the
Fortran code 
{\tt FeynHiggs}~\cite{mhiggslong,feynhiggs,mhiggsAEC,mhcMSSMlong}.
The goal for the theory precision in $\MHp$ should be the
anticipated experimental resolution or better, where an accuracy of
$\sim 1.5 \gev$ at the LHC and $\sim 0.5 \gev$ at the ILC could be
achievable (see \citere{chargedHiggs2L} for a more detailed discussion).

\begin{figure}[htb!]
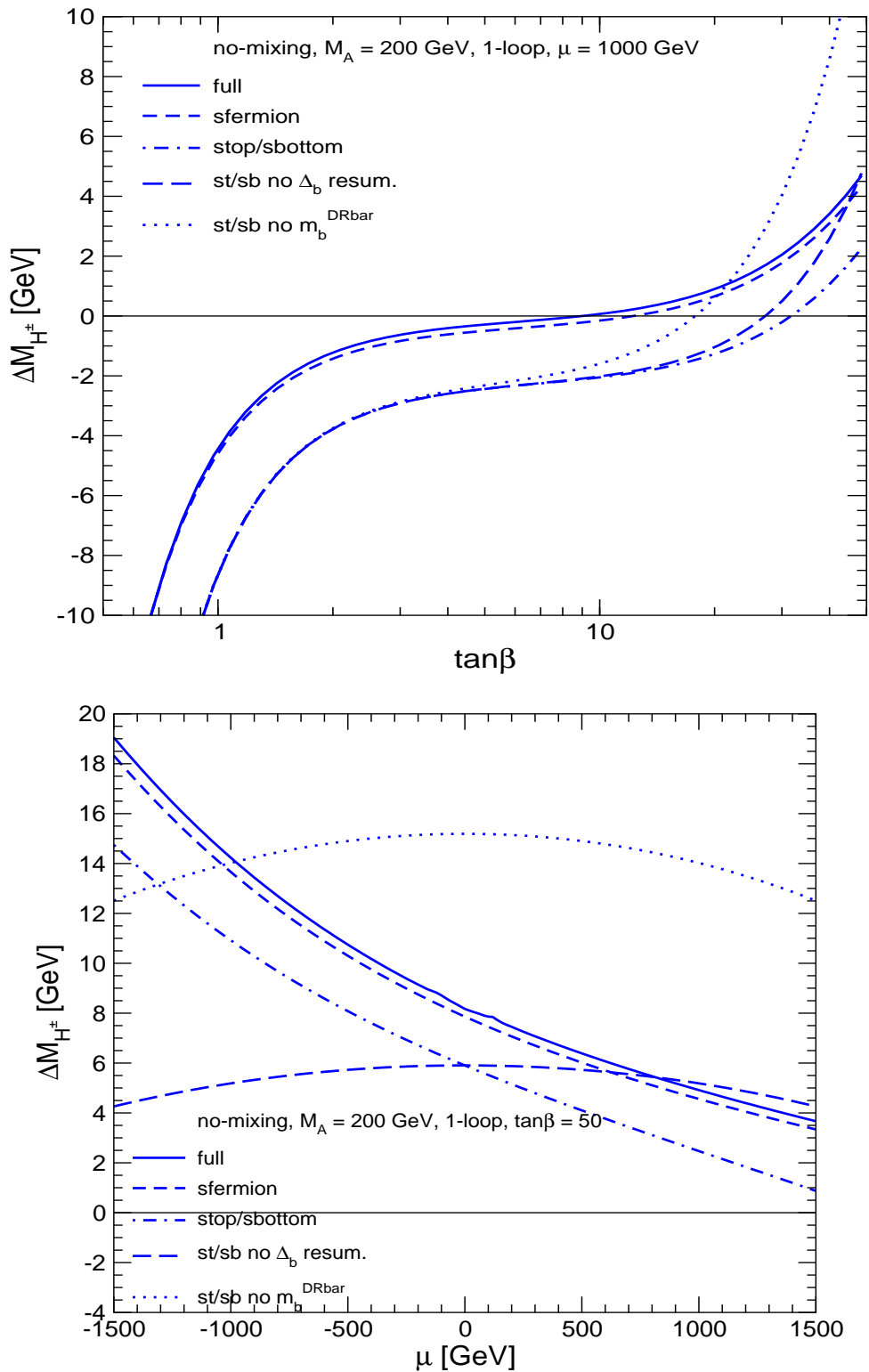

\centerline{\includegraphics[width=0.85\linewidth,height=10cm]{mhp52_cl}}
\vspace{1em}
\centerline{\includegraphics[width=0.80\linewidth,height=10cm]{mhp62_cl}}
\caption{$\De\MHp := \MHp - \mHp$ is shown as
  a function of $\MA$ (upper plot), $\tb$ (middle) and $\mu$ (lower) 
  in the $\mhmax$ scenario. $\MHp$ is evaluated at
  the two-loop (solid) and the one-loop level (dashed).
}
\label{fig:DeMHp}
\vspace{-1em}
\end{figure}

In \reffi{fig:DeMHp} we show $\De\MHp := \MHp - \mHp$
with $\MHp$ evaluated at the one-loop level in various steps of
approximation. The results are obtained in the no-mixing
scenario~\cite{benchmark2,benchmark3}.
The upper plot in \reffi{fig:DeMHp} shows $\De\MHp$ as a function of
$\tb$, while the lower plots depicts the variation with~$\mu$. 
Both parameters enter the definition of $\db$ and are thus potentially
important for a precise $\MHp$ prediction. 
In both plots the solid lines represent the full one-loop result
including the $\db$ resummation, see \refeq{def:db}. The first
approximation to this is 
shown as short-dashed line, where only the contributions from SM
fermions and their SUSY partners (i.e.\ all squarks and sleptons) are
taken into account, still including the $\db$ corrections. The next step
of approximation is shown as dot-dashed lines, where only corrections
from the $t/b$~and $\Stop/\Sbot$~sector are included, still with the
$\db$ resummation. The penultimate step of the approximation is to leave
out the $\db$ corrections, but using $\mbms$ (i.e.\ including the SM QCD 
corrections, see \refeq{mbdrbar}) in the
Higgs boson couplings, shown as the long-dashed lines. The final
step in  the approximation is to drop the SM QCD corrections, i.e.\
replacing $\mbms$ by $\mb$ in the Higgs Yukawa couplings, shown as the
dotted lines.

The dependence on $\tb$ is analyzed in the upper plot of 
\reffi{fig:DeMHp}
We show $\De\MHp$ for $\MA = 200 \gev$ and $\mu = 1000 \gev$.
The sign and size of the one-loop correction to $\MHp$ depends strongly
on $\tb$, which appears the Higgs couplings to (s)fermions as well as in
the $\db$ corrections. Negative corrections are reached for 
$\tb \lsim 10$ with $\De\MHp \approx -10 \gev$ for $\tb \approx
0.6$. It should be kept in mind that within the no-mixing
scenario values of $\tb$ around~1 are excluded by LEP Higgs
searches~\cite{LEPHiggsSM,LEPHiggsMSSM}.
Positive values of $\De\MHp$ are obtained for large $\tb$ values,
reaching $\De\MHp \approx 2 - 6 \gev$ for $\tb = 50$.
The effect of the non-sfermion
sector (short-dashed lines) is small and stays below $0.5 \gev$. 
The Yukawa coupling independent effects (dot-dashed lines) are 
$\sim 2 \gev$, largely independent of $\tb$. The contribution from the $\db$
effects is negligible for $\tb \lsim 5$ and grows with increasing $\tb$,
reaching values of up to $2 \gev$ for $\tb = 50$. 
For small values of $\mu$ (not shown here) these corrections stay very small 
even for the largest $\tb$ values.
The biggest effects on $\MHp$ can arise from the inclusion of the SM QCD
corrections to $\mb$ for $\tb \gsim 5$, reaching up to 5--$10 \gev$.

In order to analyze the dependence of the $\MHp$ prediction on $\mu$ we
show in lower plot of \reffi{fig:DeMHp}. $\De\MHp$
as a function of $\mu$ for $\MA = 200 \gev$ and $\tb = 50$.
The pure $t/b/\Stop/\Sbot$
corrections (dotted line) reach 12--$15 \gev$. Including the SM QCD
corrections (long-dashed) strongly reduced the effect to the level of 
4--$6 \gev$. In the next step the $\db$ effects are included (dot-dashed
line). Due to $\db \propto \mu\,\tb$ the inclusion of $\db$ results in a
strong asymmetry of $\MHp$ with a large correction for negative $\mu$
(corresponding to an enhanced bottom Yukawa coupling) and a much smaller
correction for positive $\mu$ (corresponding to a suppressed bottom
Yukawa coupling). $\De\MHp$ now ranges from $\sim 14 \gev$ for 
$\mu = -1500 \gev$ to $\sim 1 \gev$ for $\mu = +1500 \gev$. Adding the
corrections from the other (s)fermions yields an nearly
$\mu$-independent upward shift of $\sim 2 \gev$. Including the
non-(s)fermionic corrections result in another small upward shift of 
$\sim 0.5 \gev$. the overall one-loop effect ranges from 
$\De\MHp \approx 19 \gev$ for $\mu = -1500 \gev$ down to 
$\De\MHp \approx 3.5 \gev$ for $\mu = +1500 \gev$.

The results shown in \reffi{fig:DeMHp} clearly indicate that for a
precise $\MHp$ prediction {\em all} contributions at the one-loop level
have to be taken into account.
Effects at the GeV level may be probed at the LHC and the ILC, see above.
Thus, all parts of the one-loop calculation are relevant and should be
taken into account in a precision analysis of the charged MSSM Higgs boson.


\section{Summary}

In order to identify the Higgs sector realized in nature, the
predictions for Higgs boson masses, production cross sections and decay
widths have to be compared with experimental results.
We presented a brief overview about computer codes for the evaluation of
the properties of charged Higgs bosons. While only few codes exist for
the THDM or the NMSSM, many different tools are available for the MSSM.
These comprise tools for RGE running, the evaluation of the charged
Higgs boson mass, production cross sections and decay widths as well as
event generators. Several codes provide indirect constraints on the
charged Higgs boson sector via the evaluation of $B$-physics observables
that can be checked against existing experimental data.
We briefly compared the three codes available for the evaluation of
the mass and decay properties of the charged MSSM Higgs, boson:
{\tt FeynHiggs}, {\tt CPsuperH} and {\tt Hdecay}.
Finally, we briefly reviewed the
relevance of the various contributions to the charged MSSM Higgs boson
mass arising at the one-loop level. It was shown 
all parts of the one-loop calculation are relevant and should be
taken into account in any precision analysis involving $\MHp$.


\subsection*{Acknowledgements}

We thank the organizers of cH$^{\mbox{}^\pm}$\hspace{-2.5mm}arged 2008
for the invitation and the stimulating atmosphere.




\end{document}